\documentclass{article}

\usepackage{microtype}
\usepackage{comment}
\usepackage{graphicx}
\usepackage{subfigure}
\usepackage{booktabs} 

\usepackage{hyperref}

\usepackage[accepted]{icml2019}

\icmltitlerunning{A Hybrid Approach to Audio-to-Score Alignment}

\begin{document}

\twocolumn[
\icmltitle{A Hybrid Approach to Audio-to-Score Alignment}




\begin{icmlauthorlist}
\icmlauthor{Ruchit Agrawal}{goo}
\icmlauthor{Simon Dixon}{goo}

\end{icmlauthorlist}

\icmlaffiliation{goo}{Centre for Digital Music, Queen Mary University of London}

\icmlcorrespondingauthor{Ruchit Agrawal}{r.r.agrawal@qmul.ac.uk}

\icmlkeywords{Machine Learning, ICML}

\vskip 0.3in
]
\printAffiliationsAndNotice{This research is supported by the European Union's Horizon 2020 research and innovation programme under the Marie Skłodowska-Curie grant agreement No. 765068.\\}  

\begin{abstract}

Audio-to-score alignment aims at generating an accurate mapping between a performance audio and the score of a given piece. Standard alignment methods are based on Dynamic Time Warping (DTW) and employ handcrafted features. We explore the usage of neural networks as a preprocessing step for DTW-based automatic alignment methods. 
Experiments on music data from different acoustic conditions demonstrate that this method generates robust alignments whilst being adaptable at the same time. 

\end{abstract}
\section{Introduction and Motivation}
Audio-to-score alignment is the task of finding the optimal mapping between a performance and the score for a given piece of music. Dynamic Time Warping \cite{sakoe1978dynamic} has been the de facto standard for this task, typically incorporating handcrafted features \cite{dixon2005line, arzt2016flexible}. 
Recent advances in Music Information Retrieval have demonstrated the efficacy of Deep Neural Networks (DNNs) to a variety of tasks like music generation \cite{eck2002first}, audio classification \cite{lee2009unsupervised}, onset detection \cite{marolt2002neural}, music transcription \cite{marolt2001sonic, hawthorne2017onsets} as well as music alignment \cite{dorfer2018learning}.
The primary advantage of DNNs is that they can learn directly from data in an end-to-end manner, thereby eschewing the need for complex feature engineering. However, DNNs struggle with modelling long-term dependencies \cite{bengio1994learning} in temporal sequences. End-to-end alignment is a challenging task since it incorporates dealing with multiple inputs of different modalities, in addition to handling of very long term dependencies. This paper is an endeavor towards employing neural networks for music alignment.
We present a hybrid approach to audio-to-score alignment, which consists of a neural network based preprocessing step as a precursor to Dynamic Time Warping. This approach involves computing a frame similarity matrix which is then passed on to a DTW algorithm that computes the optimal warping path through this matrix. The advantage of our method is that the preprocessing step is trainable, thereby making our method adaptable to a particular acoustic setting, unlike traditional DTW-based methods which employ handcrafted features.

\section{Related Work}

Early works on feature learning for MIR tasks employ algorithms like Hidden Markov Models \cite{joder2013learning} or deep belief networks \cite{schmidt2012feature}. Recently, a number of works have explored feature learning for MIR using deep neural networks \cite{sigtia2014improved, oramas2017multi, thickstun2016learning, lattner2018learning, arzt2018audio, korzeniowski2016feature}. Work specifically on learning features for audio-to-score alignment has mainly focused on an evaluation of current feature representations \cite{joder2010comparative},  learning of the mapping for several common audio representations based on a best-fit criterion \cite{joder2011optimizing} and learning transposition-invariant features \cite{arzt2018audio} for alignment. \cite{hamel2013transfer} propose transfer learning for MIR tasks by  learning learn a shared latent representation across related tasks of classification and similarity detection. Weaknesses in standard approaches to choosing similarity thresholds has been explored in \cite{kinnaird2017examining}.
\cite{izmirli2010understanding} propose the idea of learning features for aligning two sequences of music, as opposed to employing a standard chroma-based feature representation.
\cite{nieto2014music} present a novel algorithm to capture music segment similarity using two-dimensional Fourier-magnitude coefficients. \cite{korzeniowski2016feature} explore frame-level audio feature learning for chord recognition using artificial neural networks. 



\section{Experiments and Results}
The standard feature representation choice for music alignment is a time-chroma representation generated from the log-frequency spectrogram. Since this representation only relies on pitch class information, it ignores variations in timbre and instrumentation, and is not adaptable to different acoustic settings.
Using neural networks helps us to override the manual feature engineering whilst providing the capability to adapt to different settings. Rather than extracting a feature representation from the inputs, we focus on the task of constructing a frame-similarity matrix. This matrix is then passed on to a DTW-based algorithm to generate the alignments. We employ a ``Siamese" Convolutional Neural Network \cite{bromley1994signature} for this task. This framework has shown promising results in the field of computer vision for computing image similarity \cite{zagoruyko2015learning}, as well as in the field of natural language processing, for learning sentence similarity \cite{mueller2016siamese} and speaker identification \cite{lukic2016speaker} amongst others. \\
We train our Siamese CNN model to determine if two patches, one from the audio and one from the synthesized MIDI ``match" or not. A similar approach has been used by \cite{izmirli2010understanding}, however they use a Multi-Layer Perceptron (MLP) framework to compute if two frames are the same or not. In addition to using an enhanced framework which is optimal for this task, our work differs from them in that we also compute similarity labels (non-binary) and use this distance matrix further for alignment. We explain the preprocessing steps below:\\ 
In order to keep the modality constant, we first convert the MIDI files to audio using FluidSynth \cite{henningsson2011fluidsynth}. We then transform the frame-level audio patches to image spectrograms using librosa \cite{mcfee2015librosa}, a Python library for audio and music analysis. We conduct experiments using both the Short-Time Fourier transform (STFT) as well as the
Constant-Q transform (CQT) transformations of the raw audios. We briefly explain our choice of loss function below:\\
The objective of our Siamese CNN model is not to classify the inputs, but to differentiate between them. Hence, a contrastive loss function is much better suited to this task than a standard classification loss function like cross entropy. The contrastive loss function is computed as follows:

\begin{figure}[ht]
\small
\vskip 0.05in
\begin{center}
\centerline{\includegraphics[width=\columnwidth]{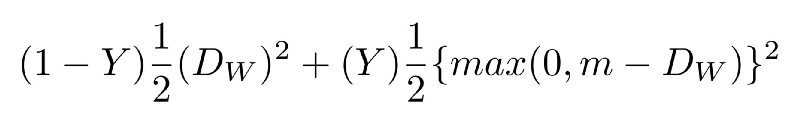}}
\label{fig:contastiveLoss}
\end{center}
\vskip -0.05in
\end{figure}
where $D_w$ is the Euclidean Distance between the outputs of the two Siamese twin networks.
More formally, $D_w$ can be expressed as follows:

\begin{figure}[ht]
\small
\vskip 0.05in
\begin{center}
\centerline{\includegraphics[width=2in]{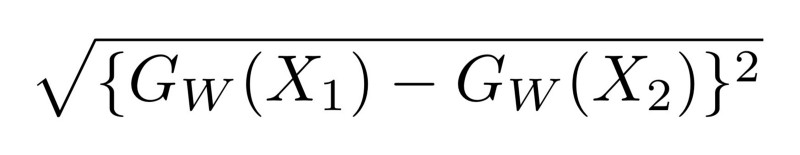}}
\label{fig:euclideanDistance}
\end{center}
\vskip -0.05in
\end{figure}
where $G_w$ is the output of each of the twin networks and $X_1$ and $X_2$ are the two inputs. 

We train the model on the MAPS database \cite{emiya2010maps}, where we have MIDI-aligned audio for a range of acoustic settings. We only select the subset containing the recordings played using a real piano, and discard the ones which are software-synthesized. 
We compute the similarity matrix using two mechanisms:
\begin{itemize}
    \item Using binary labels: For this we employ the output labels of our Siamese CNN. 0's imply similar pairs, 1's imply dissimilar pairs. 
    \item Using distances: For this we calculate $D_w$. The distance directly corresponds to the dissimilarity between the two inputs. Higher the value of $D_w$, higher the dissimilarity.
\end{itemize}
We then generate an alignment path through this matrix using fast-DTW \cite{salvador2007toward}, through a readily available DTW implementation in Python \footnote{https://pypi.org/project/fastdtw/}. We test the performance of our model on a subset of the Mazurka dataset \cite{sapp2007comparative} which contains recordings from various acoustic settings. The results obtained using both methods are given in Table \ref{results}. 
\begin{table}[H]
   \caption{Alignment accuracy (in \%)}
   \centering
   \begin{tabular}{ c c c } \toprule
      \textbf{Type of matrix} & \textbf{STFT} & \textbf{CQT}  \\ \midrule
      Binary  &  76.3 & 78.6 \\ \midrule
      Distance & 78.1  & 81.4  \\ \midrule
      \bottomrule
\end{tabular}
\label{results}
\end{table} 
Our results suggest that this method is a promising approach to alignment, especially in non-standard acoustic conditions; since the pre-processing Siamese Network is trainable on such data, unlike the manually handcrafted features used by standard DTW-based algorithms.

\section{Conclusion and Future Work}
We demonstrated that our hybrid approach to audio-to-score alignment is capable of generating robust alignments across various acoustic conditions. The advantage of our method is that it is adaptable to a particular acoustic setting without requiring a large amount of labeled training data. In the future, we would like to conduct an exhaustive evaluation of this approach on musically relevant parameters and analyze its limitations. We would also like to work on learning the features as well as the alignments in a completely end-to-end manner. 
\nocite{*}

\bibliography{example_paper}
\bibliographystyle{icml2019}


\end{document}